\documentclass[fleqn,usenatbib]{mnras}

\usepackage{newtxtext,newtxmath}
\usepackage[T1]{fontenc}

\usepackage{amsmath}
\usepackage{graphicx}
\usepackage{booktabs}
\usepackage{xcolor}

\graphicspath{{./}{figures/}}

\DeclareMathOperator{\sech}{sech}
\newcommand{\code}[1]{\texttt{#1}}

\makeatletter
\@ifclasswith{mnras}{linenumbers}{\usepackage[switch]{lineno}\linenumbers}{}
\makeatother

\title[Are $E>B$ regions essential for injection?]{\boldmath Are $E>B$ regions essential for particle injection in relativistic magnetic reconnection?}

\author[O. French et al.]{
Omar French,$^{1}$\thanks{E-mail: omar.french@colorado.edu}
Fan Guo,$^{2,3}$
Xiaocan Li,$^{2}$
Qile Zhang,$^{4}$
Yi-Hsin Liu,$^{5}$
William Daughton$^{2}$
and Hui Li$^{2}$
\\
$^{1}$Center for Integrated Plasma Studies, Department of Physics, 390 UCB, University of Colorado, Boulder, CO 80309, USA\\
$^{2}$Los Alamos National Laboratory, Los Alamos, NM 87545, USA\\
$^{3}$New Mexico Consortium, Los Alamos, NM 87544, USA\\
$^{4}$University of Maryland, College Park, MD 20742, USA\\
$^{5}$Dartmouth College, Hanover, NH 03755, USA
}

\date{Accepted XXX. Received YYY; in original form ZZZ}

\pubyear{\the\year{}}

\begin{document}
\label{firstpage}
\pagerange{\pageref{firstpage}--\pageref{lastpage}}
\maketitle

\begin{abstract}
During relativistic magnetic reconnection, X-points ($E>B$ regions at vanishing guide field) have been proposed to play an essential role in injecting particles into the nonthermal population. Using fully kinetic particle-in-cell simulations at a range of magnetizations~$\sigma \equiv B_0^2/(4\pi n_0 m_e c^2) = 12$--$800$, we study in detail how the particle energy gains up to the injection threshold~$\gamma_{\rm inj} = \sigma/4$ are partitioned into work done inside $E>B$ regions (unmagnetized), within one gyration after exit (re-magnetizing), and at all other times (magnetized). The magnetized channel supplies the largest part of the energy gain for~$\sim 95\%$ of injected tracers at~$\sigma = 12$ and~$\sim 65\%$ at~$\sigma = 800$. Most of the tracers in this magnetized majority never enter an~$E>B$ region before injection. We also compare these injection shares with the earlier studies that proposed an essential role for $E>B$ regions. We conclude that $E>B$ regions are neither necessary for injection nor the dominant direct contributors to pre-injection work by injection share in two-dimensional, nonradiative pair-plasma reconnection at vanishing guide field.
\end{abstract}


\section{Introduction} \label{sec:intro}

High-energy emission in the Universe often originates from relativistic particles with a nonthermal power-law spectrum. Relativistic magnetic reconnection, which converts magnetic energy into plasma heating and particle acceleration, has been proposed as a source of such spectra \citep[see][for recent reviews]{Guo2024,Sironi2025}. Fully kinetic particle-in-cell simulations find that, in the strongly magnetized regime~$\sigma \equiv B_0^2/(4\pi n_0 m_e c^2) > 1$ (i.e., when the magnetic energy per particle rest-mass energy is large), reconnection produces abundant nonthermal particles with a power-law distribution $f \propto \varepsilon^{-p}$, where~$\varepsilon$ is the particle kinetic energy and~$p$ approaches unity as~$\sigma$ increases \citep{Zenitani2001,Sironi2014,Guo2014,Guo2015,Werner2016}. 

In the weak guide-field regime~$b_g \equiv B_g/B_0 \ll 1$, the role of the nonideal electric field~$\mathbfit{E}_n \equiv \mathbfit{E} + \mathbfit{u} \times \mathbfit{B}/c$ (where~$\mathbfit{u}$ is the bulk plasma velocity) in producing these power laws remains under debate \citep{Zenitani2001,Guo2014,Guo2015,Guo2019,Guo2020,Sironi2022,Guo2023comment,French2023,French2026}. \citet{Zenitani2001} suggest that acceleration in electric-dominant regions (electric field larger than magnetic field, hereafter $E>B$) generates the power-law distribution, and they attribute its slope to the balance between $E>B$ acceleration and particle escape from those regions. \citet{Sironi2014} propose a similar idea that nonthermal particles originate from X-points and are then accelerated to higher energies elsewhere in the reconnection layer. These authors argue that X-points are important in establishing and determining the power law, while the rest of the reconnection layer only further extends the power law. In two-dimensional pair-plasma simulations, \citet{Nalewajko2015} identify acceleration at X-points and between merging plasmoids, while \citet{Kagan2018} attribute the spectral shape and cutoff saturation to X-point acceleration limited by secondary islands. \citet{Guo2014,Guo2015} find that a Fermi-like mechanism facilitated by particle curvature drift motion in the motional electric field~$\mathbfit{E}_m \equiv -\mathbfit{u} \times \mathbfit{B}/c$ dominates the energy gain in their simulations, and argue that a power-law distribution forms as particles are injected into and accelerated in the reconnection layer. Later analyses find that acceleration in the broader reconnection layer determines the spectrum slope and maximum energy \citep[e.g.,][]{Guo2019,Hakobyan2021}. In two-dimensional pair plasmas, continued energization by plasmoid compression allows the high-energy cutoff to grow with time \citep{Petropoulou2018}. Models also connect the spectral shape to acceleration and plasmoid trapping \citep{Uzdensky2022}, and to the coupled injection, acceleration, and escape processes \citep{Li2023}.

Following these studies, attention turned to injection, i.e., the initial energization that lifts particles from the thermal population into the power law, and to whether~$E>B$ regions provide it. \citet{Sironi2022} argues that nonideal fields control injection and are a necessary prerequisite for later acceleration. The main evidence is that particles reaching high energies have usually encountered $E>B$ or other nonideal regions. An encounter does not measure how much energy a particle gains inside those regions. Interpreting this evidence as a measure of injection therefore assumes that the encounter supplies the energy that injects the particle. \citet{Guo2023comment} show that a substantial fraction of injected particles do not pass through $E>B$ regions before injection. Particles that do enter $E>B$ regions typically stay too briefly to be injected there~\citep{Guo2023comment}. \citet{French2023} find that motional-field mechanisms such as Fermi reflection and pickup dominate the injection share in weak-guide-field reconnection. \citet{Totorica2023} make the broader claim that exact nonideal fields, not necessarily $E>B$ regions, dominate particle energization until particles first reach a threshold kinetic energy of~$\sigma m_e c^2/4$, close to the injection threshold discussed below. They base this claim on the particles with the highest final energies. This conclusion assumes that the particles reaching the highest final energies were injected in the same way as the injected population as a whole. However, particles injected by some mechanisms could reach higher energies than those injected by others, either through the injecting mechanism itself or through later acceleration, and would then be over-represented at the highest energies. \citet{Gupta2025} ask a related but narrower question by measuring the work done inside $E>B$ regions before that threshold is reached. In their analysis, the largest~$E>B$ work fractions also come from particles with the highest final energies, so the same assumption applies. \citet{French2026} extend the injection-share measurement to three dimensions, where Fermi reflection provides the largest injection share at $b_g = 0.3$. This debate has recently been summarized by \citet{Guo2025review}. Whether $E>B$ regions play an important role in injection has not been settled.

Almost all studies define a particle as injected once it surpasses an energy threshold $\gamma_{\rm inj}$, where $\gamma_{\rm inj} \simeq \sigma/4$ is reasonable for the magnetically dominated ($\sigma_h \equiv B_0^2/(4\pi h) \gg 1$, where $h$ is the plasma enthalpy density), weak-guide-field case (see \citealt{French2026} for the full expression with dependence on~$b_g$ and~$\sigma_h$). These studies differ, however, in how injection dominance is measured, and much of the apparent contradiction among earlier studies traces to that difference. The metric of \citet{French2023} and \citet{French2026}, hereafter termed \textit{injection shares}, assigns an injection mechanism to each tracer particle when it first satisfies~$\gamma > \gamma_{\rm inj}$, and reports the fraction of injected particles assigned to each mechanism. By contrast, the metrics of \citet{Sironi2022}, \citet{Totorica2023}, and \citet{Gupta2025}, hereafter termed \textit{nonthermal particle acceleration (NTPA) correlations}, condition on the energy that particles reach higher in the power-law tail. In particular, \citet{Sironi2022} asks what fraction of high-energy particles ever encountered nonideal fields. \citet{Totorica2023} and \citet{Gupta2025} ask how much of the pre-threshold work was supplied by nonideal fields or by $E>B$ regions for particles in a high-energy band. A direct comparison of these metrics in \citet{French2026} showed that they can differ substantially at $b_g = 0.3$. Particles above a high energy threshold gained most of their energy after injection, so such a selection over-represents mechanisms whose particles are injected earlier or accelerated more efficiently afterward. However, these metrics have not been compared at~$b_g = 0$, the regime where $E>B$ regions themselves are the proposed injector.

Knowing which fields inject the particles also matters for modeling reconnection acceleration in black hole magnetospheres \citep{Solanki2025} and pulsar light cylinders \citep{Soudais2024}, where global simulations must bridge the large separation between kinetic and system scales. Numerous models for particle acceleration in macroscopic reconnection layers have been developed \citep{Li2018b,Li2025,Arnold2021,Seo2024,Yin2024b,Hu2026}. If~$E>B$ regions are primarily responsible, models must account for injection in these small kinetic regions. If instead large-scale motional fields are responsible, injection is more distributed and within reach of such models \citep{Guo2019}.

In this paper, we measure the extent to which $E>B$ regions inject particles directly, through work done inside them, or potentially indirectly, by enabling energization just after a particle exits. We calculate the pre-injection work done to unmagnetized particles inside $E>B$ regions, to re-magnetizing particles after they exit those regions, and to magnetized particles at all other times. Section~\ref{sec:hypotheses} states the competing claims as testable hypotheses about where injected particles gain their energy. Section~\ref{sec:methods} describes the simulation setup. Section~\ref{sec:results} reports the contribution of each energy channel to particle injection and the correlations with high-energy particles. Section~\ref{sec:discussion} compares these results with previous work and states their limits and implications.

\section{\texorpdfstring{\boldmath The $E>B$ hypotheses}{The E>B hypotheses}} \label{sec:hypotheses}

We cast the direct and potential indirect roles of $E>B$ regions as testable hypotheses. Let us define particle~$i$ as \textit{injected} at time~$t_i^{\rm inj}$ when its Lorentz factor~$\gamma_i(t)$ first exceeds~$\gamma_{\rm inj} = \sigma/4$. This threshold corresponds to the magnetically dominated ($\sigma_h \gg 1$), zero-guide-field ($b_g=0$) limit of the injection criterion of \citet{French2026}, which requires the particle gyroradius to exceed half the thickness of the current layer at the X-point.

\begin{figure}
    \centering
    \includegraphics[width=\linewidth]{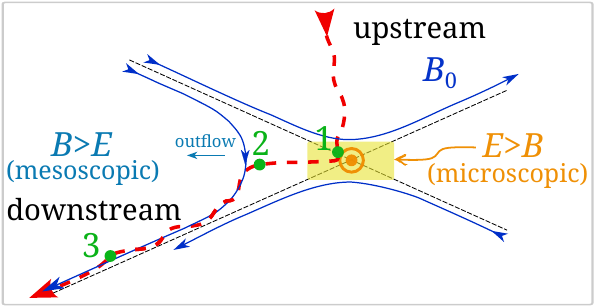}
    \caption{A particle trajectory through an X-point reconnection layer at $b_g = 0$. Yellow marks the $E>B$ region, blue curves are reconnecting magnetic field lines, and black dashed lines are separatrices. Numbered green markers distinguish three locations where the particle may gain most of its injection energy. At marker~1 the particle is unmagnetized inside $E>B$. Marker~2 lies within the post-exit re-magnetization interval of $\sim 2\pi$ radians of gyrophase. At marker~3, in the reconnection outflow well after the re-magnetization interval has ended, the particle is magnetized and can gain energy by Fermi reflection. The question of this work is which of the three locations is typical for injected particles.}
    \label{fig:cartoon}
\end{figure}

The direct $E>B$ hypothesis asserts that work delivered inside $E>B$ regions, where particles are unmagnetized, supplies most of the injection energy. This hypothesis can be tested by calculating the fraction~$\xi_i^{\rm unmag}$ of each particle's pre-injection energy gain that is delivered inside $E>B$ regions (location~1 of Fig.~\ref{fig:cartoon}). The total pre-injection work done on particle~$i$ is
\begin{equation} \label{eq:W_total}
    W_i \equiv \int_0^{t_i^{\rm inj}} dt'\, q\,\mathbfit{v}_i \cdot \mathbfit{E}_i = \left[ \gamma_i(t_i^{\rm inj}) - \gamma_i(0) \right] m_e c^2,
\end{equation}
where~$q$ is the particle charge and $\mathbfit{v}_i$ and~$\mathbfit{E}_i$ are the particle velocity and the electric field at its location. Because only the electric field does work, $W_i$ equals the particle's energy gain from the start of the simulation to its injection. We term this gain the particle's injection energy. The unmagnetized fraction is then
\begin{equation} \label{eq:xi_unmag}
    \xi_i^{\rm unmag} \equiv \frac{1}{W_i} \int_0^{t_i^{\rm inj}} \,dt' q\,\mathbfit{v}_i \cdot \mathbfit{E}_i \, \chi_i(t'),
\end{equation}
where~$\chi_i = 1$ ($=0$) while in $E>B$ ($B>E$) regions.

The indirect $E>B$ hypothesis asserts that a particle gains most of its injection energy just after it leaves an~$E>B$ region, while it re-magnetizes, so that an encounter with~$E>B$ is necessary for injection even if little work is done inside it. Each re-magnetization interval ends within~$\sim \phi$ radians of post-exit gyrophase, where~$\phi$ is of order~$2\pi$ to perhaps several~$\pi$. We adopt one full gyration, $\phi = 2\pi$, as the reference and vary~$\phi$ over~$\pi/2$--$8\pi$ in Section~\ref{sec:shares}. The work done during re-magnetization may be calculated as follows. A particle can enter and exit $E>B$ regions several times before injection. Let~$\{t_{i,k}^{\rm exit}\}_{k=1}^{K_i}$ denote the times of particle~$i$'s $K_i$ exits from~$E>B$ regions prior to injection. For each exit, let the re-magnetization interval be~$[t_{i,k}^{\rm exit}, t_{i,k}^{\rm end}]$, where~$t_{i,k}^{\rm end}$ is the earliest of (i) the time when the accumulated post-exit gyrophase first reaches~$\phi$, (ii) the next entry into~$E>B$, and (iii) injection at~$t_i^{\rm inj}$. At~$\phi = 2\pi$ and~$\sigma \geq 50$, most intervals end at the next entry into~$E>B$, before one gyration completes (Section~\ref{sec:shares}). By construction, the particle is in a $B>E$ region throughout each interval, so the relativistic gyrofrequency~$\omega_{i,\rm ce}(t) \equiv eB_i(t)/(\gamma_i(t)\,m_e c)$ is nonzero and the gyrophase keeps accumulating. The gyrophase counts elapsed time in units of the local gyroperiod and neither requires nor tests adiabatic gyration. As such, where~$B_i(t)$ is weak, the gyrophase accumulates slowly and takes longer to reach~$\phi$. The fraction of energy gained by particle~$i$ within these intervals is
\begin{equation} \label{eq:xi_remag}
    \xi_i^{\rm remag}(\phi) \equiv \frac{1}{W_i} \sum_{k=1}^{K_i} \int_{t_{i,k}^{\rm exit}}^{t_{i,k}^{\rm end}} dt'\, q\,\mathbfit{v}_i \cdot \mathbfit{E}_i.
\end{equation}

The remaining pre-injection work, done while the particle is in a $B>E$ region and outside every post-exit re-magnetization interval, defines a third, magnetized energy channel, whose fraction is given by 
\begin{equation} \label{eq:xi_mag}
    \xi_i^{\rm mag}(\phi) \equiv \frac{1}{W_i} \int_{0}^{t_i^{\rm inj}} q \mathbfit{v}_i \cdot \mathbfit{E}_i \left[ 1 - \chi_i(t') \right] \,dt' - \xi_i^{\rm remag}(\phi).
\end{equation}
This channel requires only $B>E$ at the particle, which does not guarantee adiabatic guiding-center motion near separatrices or merging plasmoids. Both $E>B$ hypotheses fall between two strengths. In their strictest form, the corresponding $E>B$-associated channel supplies nearly all of the injection energy of every injected particle ($\xi_i^{\rm unmag} \simeq 100\%$ or~$\xi_i^{\rm remag}(\phi) \simeq 100\%$). The weakest form is the least demanding claim that still makes that channel the main source of injection energy, which holds only that the channel is the largest of the three contributors to the injection energy for a majority of injected particles. A particle with~$\xi_i^{\rm mag} \simeq 100\%$ gained nearly all of its injection energy while magnetized. Such particles in substantial numbers would refute the strictest forms of both~$E>B$ hypotheses. These energy fractions quantify how much of the injection energy each channel supplies. Whether a particle encounters an $E>B$ region at all is a separate question, which Section~\ref{sec:shares} answers with the fraction of injected particles that never enter one before injection.

Figure~\ref{fig:cartoon} illustrates how a particle realizes the direct hypothesis by gaining most of its injection energy at location~1, the indirect hypothesis at location~2, and neither at location~3.

\section{Simulation setup} \label{sec:methods}

To measure each channel's fraction of the pre-injection work, we perform two-dimensional, collisionless, fully kinetic particle-in-cell simulations with the \code{VPIC} code, which solves the relativistic Vlasov--Maxwell equations \citep{Bowers2008}.

All runs start from a force-free current sheet in an electron-positron plasma. The density is normalized by the uniform upstream total pair density~$n_0 = n_e + n_i$. Each cell initially contains~$100$ electron-positron pairs (i.e., of mass ratio~$m_i/m_e = 1$) of temperature~$k_B T_e = k_B T_i = 0.36\,m_e c^2$. The cold upstream magnetization is~$\sigma \equiv B_0^2 / (4 \pi n_0 m_e c^2)$, where~$B_0$ is the asymptotic reconnecting magnetic field. We analyze nine runs (Table~\ref{tab:sample_counts}) along two axes. A magnetization scan varies~$\sigma \in \{12, 50, 200, 800\}$ at fixed box size~$L_x/d_e = 1600$, where~$d_e \equiv c/\omega_{\rm pe}$ is the non-relativistic electron skin depth and~$\omega_{\rm pe} \equiv \sqrt{4 \pi n_e e^2 / m_e}$ is the electron plasma frequency, with~$n_e = n_0/2$ the upstream electron density. A box-size scan varies~$L_x/d_e \in \{200, 400, 800, 1600, 3200, 6400\}$ at fixed~$\sigma = 50$. The~$\sigma = 50$, $L_x/d_e = 1600$ run belongs to both scans.

Each domain spans~$x \in [0, L_x]$ and~$z \in [-L_z/2, +L_z/2]$, with~$L_z = 0.625\,L_x$. Every run has resolution~$\Delta x = \Delta y = \Delta z = 0.25\,d_e$ and one cell along~$y$. The simulations retain all three field and momentum components but resolve spatial variation only in the~$x$--$z$ plane. Periodic boundary conditions are set for fields and particles in the~$x$-direction. Conducting boundaries are set for fields and reflecting boundaries are set for particles in the~$z$-direction. The timestep is~$\omega_{\rm pe}\Delta t = 0.16$. Runs evolve for~$\simeq 4000\,\omega_{\rm pe}^{-1}$, except the~$L_x/d_e = 6400$ run, which reaches~$\simeq 7600\,\omega_{\rm pe}^{-1} \simeq 1.2\,L_x/c$. All measurements below lie within these intervals. We apply no additional filtering to the fields or currents.

\begin{table}
    \setlength{\tabcolsep}{3pt}
    \centering
    \caption{Sample sizes for the nine simulations, each tracking~$6.4 \times 10^5$ tracer particles. The injected column gives the fraction of all tracked tracers that are injected between the reconnection onset~$t_{\rm onset}$ and~$t_{\rm onset}+L_x/c$, the population of Sections~\ref{sec:raw_distributions} and~\ref{sec:shares}. The high-energy column gives the fraction of those injected tracers with kinetic energy~$\gamma_i - 1 \geq \sigma$ at~$t_{\rm onset}+L_x/c$, the high-energy sample of Section~\ref{sec:ntpa}.}
    \label{tab:sample_counts}
    \begin{tabular*}{\columnwidth}{@{\extracolsep{\fill}}rrrr@{}}
        \toprule
        $\sigma$ & $L_x/d_e$ & Injected & High-energy \\
        \midrule
        12 & 1600 & 29\% & 12\% \\
        50 & 200 & 6.1\% & 11\% \\
        50 & 400 & 15\% & 12\% \\
        50 & 800 & 17\% & 15\% \\
        50 & 1600 & 19\% & 19\% \\
        50 & 3200 & 19\% & 20\% \\
        50 & 6400 & 25\% & 20\% \\
        200 & 1600 & 12\% & 22\% \\
        800 & 1600 & 6.6\% & 26\% \\
        \bottomrule
    \end{tabular*}
\end{table}

The initial magnetic field is
\begin{equation} \label{eq:initial_B}
   \mathbfit{B} = B_0 \tanh{(z / \delta)} \, \hat{\mathbfit{x}} + B_0 \sqrt{\sech^2(z / \delta) + b_g^2} \, \hat{\mathbfit{y}},
\end{equation}
where~$\delta$ is the initial sheet half-thickness and~$b_g \equiv B_g / B_0$ is the asymptotic guide-field strength. Normal to the reconnection plane, the~$\hat{\mathbfit{y}}$ component combines the localized field~$B_0\,\sech(z/\delta)$ in the sheet with the uniform guide field~$B_g$. We set the half-thickness to~$3\,d_e$ at~$\sigma = 12$, $6\,d_e$ at~$\sigma = 50$, $12\,d_e$ at~$\sigma = 200$, and~$24\,d_e$ at~$\sigma = 800$. The thicker sheets ensure that the initial current density can be achieved by the drift velocity of the available charge carriers ($J = cB_0/(4\pi\delta) = n_0 e v_d < n_0 e c$). Every run has a vanishing guide-field strength~$b_g = 10^{-5}$. We trigger reconnection in every run with a small initial perturbation of the form used by \citet{Birn2001} and \citet{Kilian2020}, and we retain the initial current-carrying particles, which \citet{Kilian2020} conclude have a negligible influence on the resulting particle acceleration. To analyze particle injection, each simulation initially selects~$6.4 \times 10^5$ electrons uniformly at random and tracks their positions, momenta, and the local electric and magnetic fields at every simulation timestep. We exclude radiative losses.

Times are measured relative to the reconnection onset~$t_{\rm onset}$, the first appearance of an $E>B$ region, which we detect as the earliest $E>B$ sample along any tracer trajectory. Tracers injected before~$t_{\rm onset}$ originate in the initial current sheet and reach~$\gamma_{\rm inj}$ without $E>B$ contact, because no $E>B$ region yet exists. We exclude them as initialization artifacts, which raises the fraction of injected tracers with $E>B$ contact and so favors the $E>B$ hypotheses. This pre-onset cohort grows from~$\sim 1\%$ of all tracers injected during the run at~$\sigma = 50$ to~$\sim 15\%$ at~$\sigma = 800$. Unless noted otherwise, we analyze only the tracers injected between~$t_{\rm onset}$ and~$t_{\rm onset}+L_x/c$, one light-crossing time after onset, by which time the injection shares have stabilized at~$\sigma = 50$ (Section~\ref{sec:shares}). Table~\ref{tab:sample_counts} gives the fraction of tracers in this population for each run.

We record injection at the first timestep with~$\gamma_i>\gamma_{\rm inj}$ and sum channel work as~$q\,\mathbfit{v}_i\cdot\mathbfit{E}_i\,\Delta t$ through that timestep, without interpolating for the precise threshold or $E>B$ boundary crossings. The post-exit gyrophase is the time integral of the relativistic gyrofrequency~$\omega_{i,\rm ce}$, which we evaluate with the trapezoidal rule over successive timesteps. We divide each channel's work by the sum of the three channel works, so the three fractions sum to one by construction. This normalization does not test whether the summed work equals the particle's energy change. The summed work usually has excellent agreement with the actual particle energy change.

\section{Results} \label{sec:results}

\begin{figure*}
    \centering
    \includegraphics[width=\textwidth]{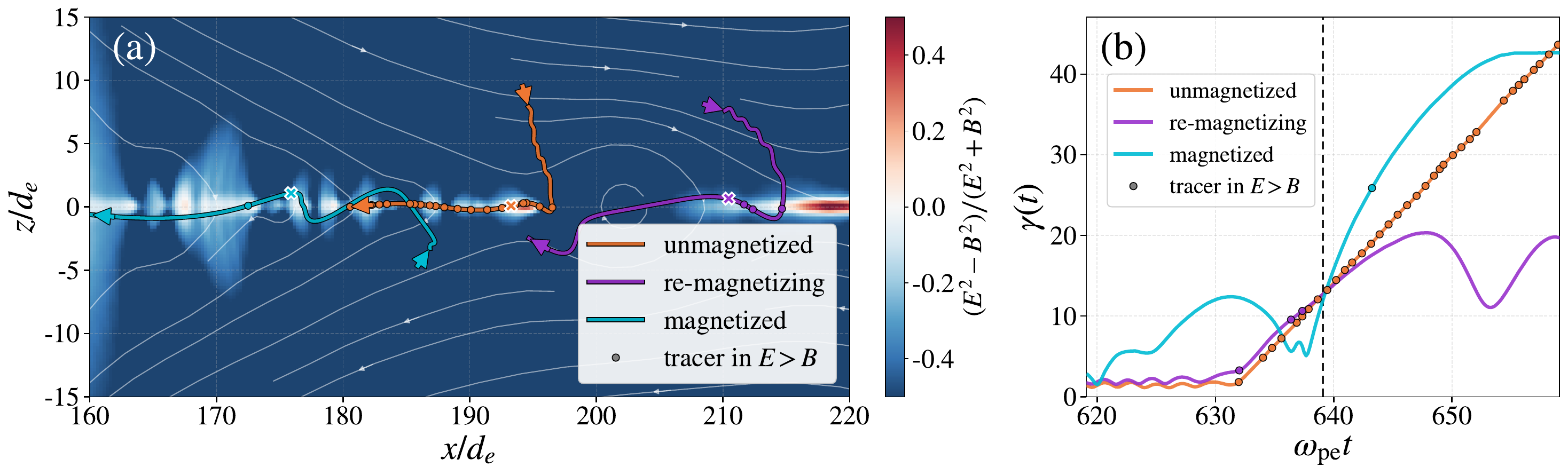}
    \caption{
    Three example injected tracer particles from the $\sigma = 50$, $L_x/d_e = 1600$ run, each of which realizes one energy channel of Section~\ref{sec:hypotheses} (unmagnetized in orange, re-magnetizing in purple, magnetized in cyan). Panel~(a) shows their trajectories in the reconnection plane, with evenly spaced dots where the particle is inside $E>B$ (at least one per contact, even a single timestep) and with arrowheads giving the direction of motion. Crosses mark where each tracer first reaches $\gamma_{\rm inj}$. The background shows the electric dominance $(E^2 - B^2)/(E^2 + B^2)$ in a single field snapshot at~$\omega_{\rm pe}t \simeq 639$. Red regions have $E>B$, blue regions have~$B>E$, and white streamlines trace the in-plane magnetic field. Panel~(b) shows the corresponding Lorentz-factor histories, with the same colors and dots, from~$20\,\omega_{\rm pe}^{-1}$ before to~$20\,\omega_{\rm pe}^{-1}$ after injection. All three tracers reach~$\gamma_{\rm inj}$ within~$\sim 0.3\,\omega_{\rm pe}^{-1}$ of one another, at~$\omega_{\rm pe}t \simeq 639$ (vertical dashed line). The orange tracer reaches $\gamma_{\rm inj}$ inside an $E>B$ region and the purple tracer just after exiting one. The cyan tracer never enters~$E>B$ before injection, and its one dot in each panel marks a brief contact after injection.
    }
    \label{fig:example_trajectories}
\end{figure*}

We begin with examples of particles injected by each energy channel defined in Section~\ref{sec:hypotheses}. Figure~\ref{fig:example_trajectories} shows three such tracers from the~$\sigma = 50$, $L_x/d_e = 1600$ run, one per channel. Panel~(a) shows their trajectories near an $E>B$ region, and panel~(b) shows their Lorentz-factor histories. The orange tracer gains most of its injection energy inside the region. The purple tracer is injected soon after exiting it. The cyan tracer is injected while magnetized, without entering $E>B$ before injection.

The rest of this section measures how often each case occurs across the injected population. Section~\ref{sec:raw_distributions} presents the distributions of the unmagnetized, re-magnetizing, and magnetized energy fractions, $\xi_i^{\rm unmag}$, $\xi_i^{\rm remag}$, and $\xi_i^{\rm mag}$. Section~\ref{sec:shares} measures the contribution of each channel to the injected-particle population. Section~\ref{sec:ntpa} asks how those same populations are represented among high-energy particles.

\subsection{Per-particle energy-fraction distributions} \label{sec:raw_distributions}

\begin{figure}
    \centering
    \includegraphics[width=\linewidth]{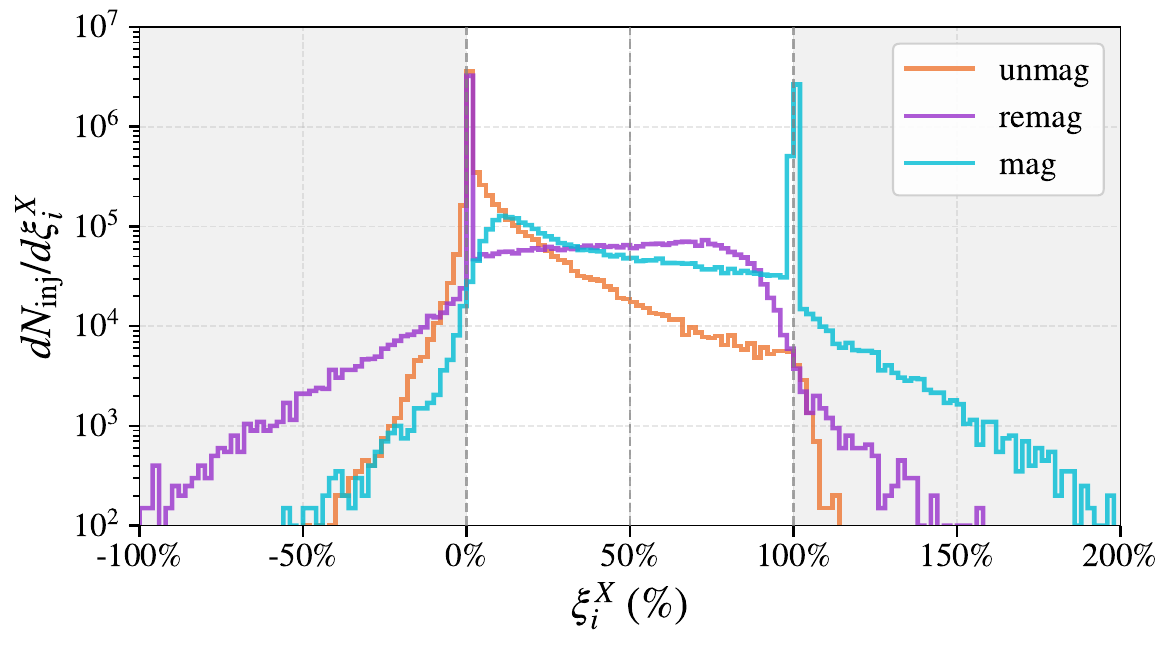}
    \caption{
    Distributions of the unmagnetized (orange, equation~\eqref{eq:xi_unmag}), re-magnetizing (purple, equation~\eqref{eq:xi_remag}), and magnetized (cyan, equation~\eqref{eq:xi_mag}) energy fractions at~$\sigma = 50$, $L_x/d_e = 1600$, for tracers injected within one light-crossing time of reconnection onset. Re-magnetization intervals end at most one gyration (i.e., $\phi = 2\pi$ radians of gyrophase) after exit from~$E>B$ (Section~\ref{sec:hypotheses}). Gray shading marks~$\xi_i^X<0$ and~$\xi_i^X>1$. Vertical dashed lines mark $0\%$, $50\%$, and~$100\%$. The magnetized distribution peaks near~$100\%$, whereas the unmagnetized and re-magnetizing distributions peak near~$0\%$.
    }
    \label{fig:raw_distributions}
\end{figure}

For each particle~$i$, let~$\xi_i^X$ denote the fraction of the injection energy supplied while in state~$X$, where~$X$ can be unmagnetized, re-magnetizing, or magnetized. These fractions of all injected particles form distributions~$dN_{\rm inj}/d\xi_i^X$ shown in Fig.~\ref{fig:raw_distributions}, where~$N_{\rm inj}$ is the number of injected tracers. The post-exit re-magnetization intervals use~$\phi = 2\pi$ (i.e., one gyration). Each tracer's three fractions sum to one. A fraction can still be negative when the relevant mechanism does net negative work, and it exceeds unity when the other two sum to a negative value.

The unmagnetized and re-magnetizing distributions peak near zero contribution, whereas the magnetized distribution peaks near unity. Almost every injected tracer receives some magnetized work before crossing $\gamma_{\rm inj}$, and for most tracers that work is the largest contribution. $E>B$-associated work (unmagnetized or re-magnetizing) contributes to many injection histories, but it rarely supplies nearly all of the injection energy.

The signed tails in Fig.~\ref{fig:raw_distributions} show that the net work in each channel can also be negative. Many particles exit an~$E>B$ region with the local electric force opposing their motion and are initially decelerated, producing the $\xi_i^{\rm remag}<0$ tail. Magnetized work usually offsets those losses (most tracers in that tail have $\xi_i^{\rm mag}>1$). The converse, in which re-magnetizing gains offset magnetized losses, occurs less often. At most~$\sim 1.2\%$ of tracers exceed~$\xi_i^X = 1$ in either $E>B$-associated channel at any magnetization in the scan.

At~$\sigma \geq 50$, the re-magnetizing channel is also built from many short intervals rather than a single clean post-exit interval. An injected tracer that enters~$E>B$ before injection may do so repeatedly. Among such tracers, the median number of pre-injection exits rises from~$1$ at~$\sigma = 12$ to~$8$ at~$\sigma \geq 200$, and each stay is brief. The mean residence time per encounter rises from~$\sim 0.4$ to~$\sim 3\,\omega_{\rm pe}^{-1}$ across the scan, far below the~$\gtrsim 12.5\,\omega_{\rm pe}^{-1}$ that direct acceleration by the reconnection electric field needs to reach~$\gamma_{\rm inj}$ at~$\sigma = 50$ \citep{Guo2023comment}. These repeated short stays suggest that many particles skim the boundaries of~$E>B$ regions. Each exit opens a new post-exit interval. Section~\ref{sec:shares} returns to how this interval structure fragments with magnetization.

\subsection{Injection shares of each population} \label{sec:shares}

\begin{figure*}
    \centering
    \includegraphics[width=\textwidth]{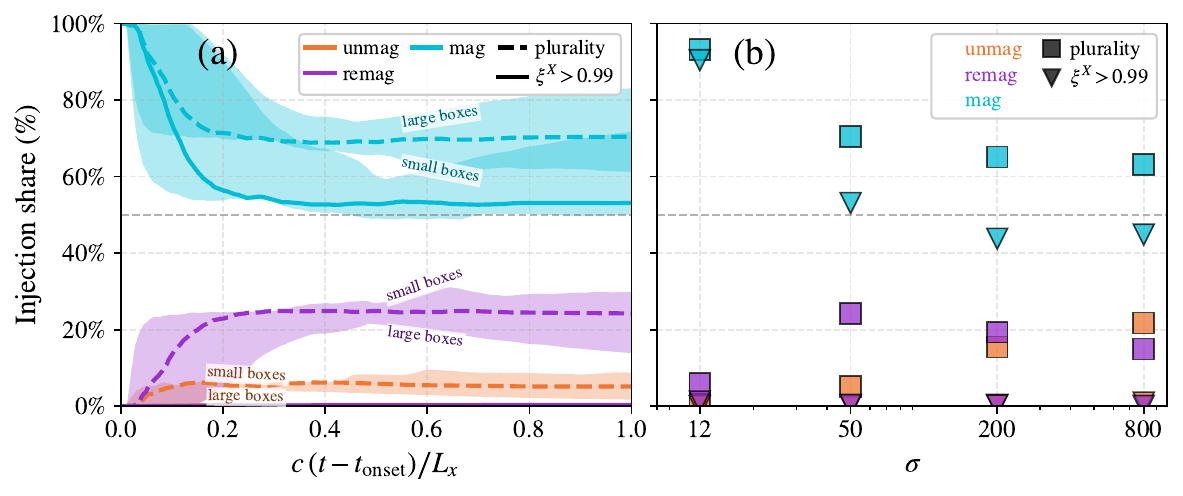}
    \caption{
    Injection shares for the unmagnetized (orange), re-magnetizing (purple), and magnetized (cyan) energy channels. Panel~(a) shows cumulative shares at~$\sigma = 50$, $L_x/d_e = 1600$ over time elapsed since reconnection onset, where dashed lines use the plurality assignment and solid lines use the pure assignment~$\xi_i^X > 99\%$. Shaded bands give the minimum and maximum shares across~$L_x/d_e \in \{200, 400, 800, 1600, 3200, 6400\}$ at~$\sigma = 50$, with one band per channel for each assignment rule. Only the magnetized channel therefore shows two distinct bands, because the pure bands of the other two channels lie near zero. Band edges are labeled by whether a small ($L_x/d_e \leq 800$) or large ($L_x/d_e \geq 3200$) box forms them. Panel~(b) shows the shares one light-crossing time after reconnection onset, where squares use plurality assignment and downward triangles use pure assignment. Pure assignment leaves some tracers unassigned, so the three pure shares sum to less than~$100\%$.
    }
    \label{fig:shares}
\end{figure*}

We now assign each injected tracer to a channel by its energy fractions and measure each channel's injection share, the fraction of injected tracers assigned to it. We consider two assignment rules.
\begin{enumerate}
    \item \textit{Plurality assignment.} A tracer is assigned to the channel with the largest energy fraction.
    \item \textit{Pure assignment.} A tracer is assigned to channel~$X$ only if~$\xi_i^X > 99\%$ and no other channel also exceeds~$99\%$. All other tracers remain unassigned.
\end{enumerate}
Figure~\ref{fig:shares} shows how the injection shares evolve in time and vary with box size and magnetization, and Table~\ref{tab:channel_populations} lists the values across the magnetization scan.

\begin{table}
    \centering
    \caption{Injection shares and negligible-work fractions of the three energy channels, and $E>B$ contact before injection, across the magnetization scan~$\sigma \in \{12, 50, 200, 800\}$, using $\gamma_{\rm inj}=\sigma/4$ and~$\phi=2\pi$. All entries are computed from tracers injected between~$t_{\rm onset}$ and~$t_{\rm onset} + L_x/c$, the same population as Fig.~\ref{fig:shares}. The plurality and pure blocks list the values plotted in panel~(b) of Fig.~\ref{fig:shares}. Plurality assigns each tracer to the channel with the largest energy fraction. Pure assignment requires exactly one channel to satisfy~$\xi_i^X > 99\%$. Some tracers remain unassigned, so the three pure shares sum to less than~$100\%$. The negligible block instead applies~$\xi_i^X < 1\%$ independently to each channel, including negative net work and small positive contributions. In the last block, which is not resolved by channel, the first row gives the fraction of injected tracers that never enter an $E>B$ region before injection. Instead of tracers, the second row counts the pre-injection post-exit intervals of all injected tracers and gives the fraction cut short by the next $E>B$ entry, before~$\phi$ radians of gyrophase accumulate or injection occurs. The pre-onset cohort excluded in Section~\ref{sec:methods}, which is magnetized by construction, would make up~$\sim 30\%$ of all tracers injected by~$t_{\rm onset} + L_x/c$ at~$\sigma = 800$. Including it raises the magnetized plurality share there to~$\sim 75\%$.
    }
    \label{tab:channel_populations}
    \begin{tabular}{lcccc}
        \toprule
        Channel & $\sigma = 12$ & $\sigma = 50$ & $\sigma = 200$ & $\sigma = 800$ \\
        \midrule
        \multicolumn{5}{l}{\textit{Plurality (largest channel)}} \\
        Unmagnetized & $0.53\%$ & $5.2\%$ & $15\%$ & $22\%$ \\
        Re-magnetizing & $6.0\%$ & $24\%$ & $19\%$ & $15\%$ \\
        Magnetized & $93\%$ & $70\%$ & $65\%$ & $63\%$ \\
        \midrule
        \multicolumn{5}{l}{\textit{Pure ($\xi_i^X > 99\%$)}} \\
        Unmagnetized & $0.064\%$ & $0.21\%$ & $0.37\%$ & $0.98\%$ \\
        Re-magnetizing & $1.3\%$ & $0.36\%$ & $0.25\%$ & $0.11\%$ \\
        Magnetized & $90\%$ & $53\%$ & $44\%$ & $45\%$ \\
        \midrule
        \multicolumn{5}{l}{\textit{Negligible ($\xi_i^X < 1\%$)}} \\
        Unmagnetized & $93\%$ & $59\%$ & $45\%$ & $46\%$ \\
        Re-magnetizing & $91\%$ & $56\%$ & $51\%$ & $54\%$ \\
        Magnetized & $2.1\%$ & $0.95\%$ & $1.1\%$ & $2.0\%$ \\
        \midrule
        \multicolumn{5}{l}{\textit{$E>B$ contact before injection}} \\
        No contact & $90\%$ & $50\%$ & $39\%$ & $43\%$ \\
        Cut by re-entry & $49\%$ & $79\%$ & $92\%$ & $94\%$ \\
        \bottomrule
    \end{tabular}
\end{table}

Panel~(a) of Fig.~\ref{fig:shares} shows the cumulative shares at~$\sigma = 50$, $L_x/d_e = 1600$ during the first light-crossing time after reconnection onset. The shares stabilize within~$t - t_{\rm onset} \sim 0.25\,L_x/c$, and their values at one light-crossing time form the~$\sigma = 50$ column of Table~\ref{tab:channel_populations}. Across the box-size scan, larger boxes give a larger magnetized plurality share and a smaller unmagnetized one (shaded bands in panel~(a)), although the~$L_x/d_e = 400$ and~$800$ boxes give nearly identical shares. The finite box therefore biases the measurement toward the~$E>B$-associated channels.

Panel~(b) shows the shares at one light-crossing time after onset across the magnetization scan. The magnetized channel holds the largest plurality share at every magnetization (Table~\ref{tab:channel_populations}). From~$\sigma = 12$ to~$\sigma = 800$, the unmagnetized plurality share grows by a smaller factor with each fourfold step in~$\sigma$ and reaches only~$\sim 20\%$. Because each tracer's three fractions sum to one, plurality assignment can credit a channel that supplied only about a third of the injection energy, and even this generous rule leaves the unmagnetized share well below half. The re-magnetizing plurality share peaks at~$\sigma = 50$ and declines toward higher magnetization.

The contrast between the magnetized and $E>B$-associated channels is sharpened by the ``Pure'' and ``Negligible'' blocks of Table~\ref{tab:channel_populations}. Under pure assignment, almost no injected tracers at any magnetization are assigned to either $E>B$-associated channel. Conversely, almost no tracers have a magnetized work fraction below~$1\%$, while roughly half or more fall below~$1\%$ in the unmagnetized or re-magnetizing channel at every magnetization. The magnetized pure share is large partly by construction. A tracer that never enters~$E>B$ before injection has zero unmagnetized and re-magnetizing fractions, so it counts as pure magnetized and as negligible in both $E>B$-associated channels. The magnetized pure share exceeds the fraction of such tracers (\textit{No contact} row of the table) by at most~$\sim 5$ percentage points, so only a few percent of injected tracers enter~$E>B$ and still gain more than~$99\%$ of their injection energy while magnetized.

The unmagnetized and re-magnetizing plurality shares have opposite trends, while their sum grows by~$\sim 8$ percentage points from~$\sigma = 50$ to~$\sigma = 800$. Over the same range, the fraction of pre-injection post-exit intervals that end at the next~$E>B$ entry rises from~$\sim80\%$ to~$\sim95\%$ (Table~\ref{tab:channel_populations}). Re-entry terminates an interval by definition, and work during the subsequent~$E>B$ residence belongs to the unmagnetized channel. These trends are consistent with more frequent interruptions of post-exit energization, but do not establish that the same work is transferred from the re-magnetizing to the unmagnetized channel across different simulations. Varying the post-exit accumulated gyrophase~$\phi$ over~$\pi/2$--$8\pi$ raises the re-magnetizing plurality share to at most~$\sim 30\%$, and the magnetized channel retains the largest plurality share for every combination of~$\sigma$ and~$\phi$.

The fraction of tracers injected without once entering~$E>B$ falls from~$\sim 90\%$ at~$\sigma = 12$ to~$\sim 40$--$45\%$ at~$\sigma \geq 200$ (Table~\ref{tab:channel_populations}). Since these particles are injected only after~$E>B$ regions appear, their injection proceeds while such regions exist elsewhere in the domain yet involves no~$E>B$ contact at all. Here we read~$E>B$ contact in its widest form, counting even a single tracer timestep inside a region. Requiring at least two consecutive timesteps instead raises the fraction injected without~$E>B$ contact at~$\sigma = 50$ from~$\sim 50\%$ to~$\sim 70\%$, so for~$\sim 20\%$ of injected tracers the entire pre-injection~$E>B$ contact consists of single-step crossings. 

As an initialization check, we additionally exclude tracers initially within~$|z|<2\delta$, retaining the population cut of Section~\ref{sec:methods}. Plurality shares then shift by at most~$\sim 5$ percentage points, pure shares by at most~$\sim 7$, and the fractions injected without entering an~$E>B$ region by at most~$\sim 8$. The magnetized channel retains the largest injection share under both assignment rules at every magnetization.

\subsection{\texorpdfstring{Correlations between injection channels and post-injection energization}{Correlations between injection channels and post-injection energization}} \label{sec:ntpa}

\begin{figure*}
    \centering
    \includegraphics[width=\textwidth]{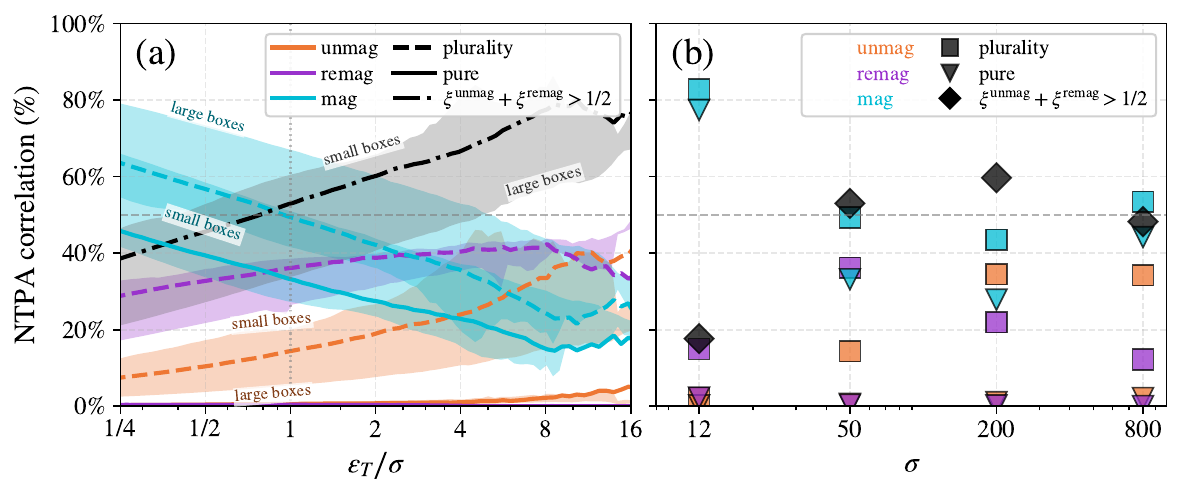}
    \caption{
    NTPA correlations for the particles injected while unmagnetized (orange), re-magnetizing (purple), and magnetized (cyan). The selected tracers are restricted to those injected by one light-crossing time after reconnection onset and conditioned on their kinetic energy at that time satisfying $\gamma_i - 1 \geq \varepsilon_T$. Panel~(a) shows how each NTPA correlation varies with increasing energy threshold in the case of~$\sigma = 50$, $L_x/d_e = 1600$, for plurality assignment (dashed) and pure assignment (solid). The black dash-dot curve in panel~(a) and black diamonds in panel~(b) show the combined $E>B$-associated fraction. This criterion counts tracers with~$\xi_i^{\rm unmag}+\xi_i^{\rm remag}>1/2$ rather than adding the two channel populations. Shaded bands give the minimum and maximum NTPA correlations across~$L_x/d_e \in \{200, 400, 800, 1600, 3200, 6400\}$ at~$\sigma = 50$, with one band in each curve's color for each assignment rule and a gray band for the combined criterion. Band edges are labeled by whether a small ($L_x/d_e \leq 800$) or large ($L_x/d_e \geq 3200$) box forms them. The re-magnetizing band is not ordered by box size and carries no labels. A run contributes to the bands only where it holds at least~$100$ selected tracers, and the~$L_x/d_e = 200$ run only up to~$\varepsilon_T = \sigma$, as in Fig.~\ref{fig:ntpa_boxsize}. Under plurality assignment, the unmagnetized and re-magnetizing NTPA correlations rise with threshold, whereas under pure assignment they stay near zero. Panel~(b) shows each NTPA correlation at the fixed threshold~$\varepsilon_T = \sigma$ across the magnetization scan, for plurality assignment (squares) and pure assignment (downward triangles). As with Fig.~\ref{fig:shares}, a horizontal dashed line marks $50\%$. The unassigned pure fraction is one minus the sum of the three mutually exclusive pure fractions and is not plotted.
    }
    \label{fig:ntpa_correlations}
\end{figure*}

Let us now ask which injected-particle populations appear among particles above a chosen kinetic energy. Such a census is useful for comparison with previous measurements of \citet{Sironi2022}, \citet{Totorica2023}, \citet{Gupta2025}, and \citet{French2026}. As in Section~\ref{sec:intro}, we term these \textit{NTPA correlations}, following \citet{French2026}. As with the injection shares in Section~\ref{sec:shares}, we evaluate the NTPA correlations one light-crossing time after reconnection onset, by which time the shares have stabilized (Fig.~\ref{fig:shares}). For each kinetic-energy threshold~$\varepsilon_T$, in units of~$m_e c^2$, we select particles with~$\gamma_i - 1 \geq \varepsilon_T$ and measure what fraction of that selected set belongs to each channel. The NTPA correlations are shown in Fig.~\ref{fig:ntpa_correlations}. 

Panel~(a) illustrates the consequence of high-energy selection at~$\sigma = 50$, $L_x/d_e = 1600$. Under plurality assignment, the unmagnetized and re-magnetizing NTPA correlations rise as the energy threshold increases, while the magnetized NTPA correlation falls. Under pure assignment, the unmagnetized and re-magnetizing NTPA correlations remain close to zero across the plotted energy range, while the magnetized NTPA correlation remains substantial. The black dash-dot curve shows the combined $E>B$-associated fraction, the fraction of selected tracers whose combined unmagnetized and re-magnetizing energy fractions satisfy~$\xi_i^{\rm unmag} + \xi_i^{\rm remag} > 1/2$. This criterion sums the work from both stages. The combined $E>B$-associated fraction rises from below half near the injection threshold to a large majority at the highest thresholds shown. High-energy selection can therefore inflate the $E>B$-associated fractions under the plurality rule and the combined criterion. Even at the highest thresholds shown, however, only a few percent of the selected tracers gained nearly all of their injection energy while unmagnetized or re-magnetizing (the pure-assignment curves).

Panel~(b) shows NTPA correlations at the fixed threshold~$\varepsilon_T = \sigma$ for upstream magnetizations~$\sigma \in \{ 12, 50, 200, 800\}$. The unmagnetized NTPA correlation rises with magnetization to~$\sim 35\%$ at~$\sigma = 200$--$800$, well above the corresponding injection shares in Fig.~\ref{fig:shares}. At~$\sigma = 50$, tracers above the threshold were injected at a median of~$\sim 0.3\,L_x/c$ after onset, against~$\sim 0.5\,L_x/c$ for the entire injected population. The census therefore favors early injection, because tracers injected earlier have more time to gain energy before the census is taken. \citet{French2026} call this weighting the head-start effect. We next ask whether injection timing also explains why the unmagnetized NTPA correlation exceeds the unmagnetized injection share. To separate timing from later energy gain, we split the injected tracers into ten bins of injection time, each~$0.1\,L_x/c$ wide, and compare each channel's injection share with its NTPA correlation bin by bin. The three populations are injected over largely overlapping times, so differences in injection time add at most~$\sim 1$ percentage point to that excess at~$\sigma \leq 200$, and at~$\sigma = 800$, where timing favors the magnetized channel, they reduce it by~$\sim 3$ points. In every bin at~$\sigma = 50$, a tracer injected while unmagnetized is instead~$\sim 2$--$9$ times more likely to be above the threshold at the census than one injected while magnetized. The excess therefore arises after injection. A high-energy census weights each channel by the energy its tracers gain afterward, which differs by channel, and so does not measure the channel's injection share. We have not identified the mechanism behind this difference in post-injection energy gain. Even with this excess, the magnetized NTPA correlation stays the largest at every magnetization (Fig.~\ref{fig:ntpa_correlations}(b)). Including the pre-onset cohort instead raises the~$\sigma = 800$ magnetized NTPA correlation to~$\sim 75\%$, as that cohort, injected before onset, has the most time for later acceleration. Either way, the unmagnetized channel remains at or below~$\sim 35\%$ of the high-energy census at every magnetization in the scan. At the stricter threshold~$\varepsilon_T = 4\sigma$ it reaches~$\sim 50\%$ at~$\sigma = 200$, still at or below about half. The combined $E>B$-associated fraction peaks at~$\sim 60\%$ at~$\sigma = 200$ and falls at~$\sigma = 800$. That drop comes from the re-magnetizing NTPA correlation, which falls steeply above~$\sigma = 50$ (Fig.~\ref{fig:ntpa_correlations}(b)) while the unmagnetized NTPA correlation grows. As with the injection shares, these opposite trends are consistent with post-exit intervals ending more often at re-entry into~$E>B$ (Section~\ref{sec:shares}). The opposite trends do not by themselves identify the fields responsible for post-exit energization (Section~\ref{sec:discussion}).

\begin{figure}
    \centering
    \includegraphics[width=\linewidth]{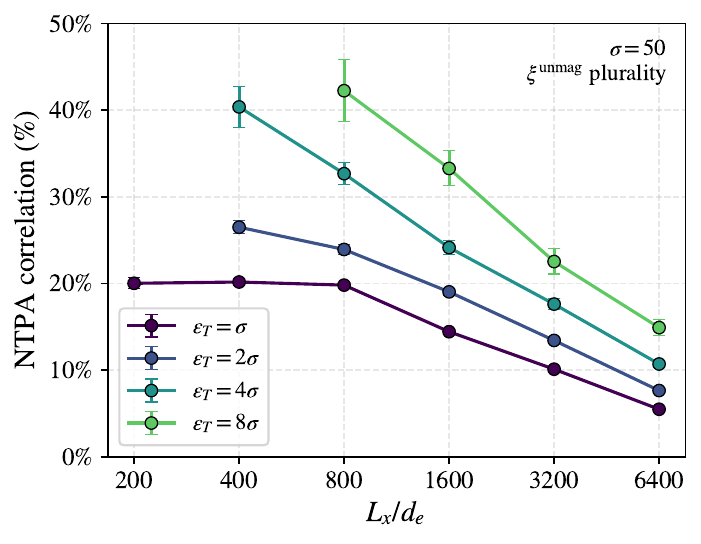}
    \caption{
    NTPA correlation of the unmagnetized population against box size~$L_x/d_e$ at~$\sigma = 50$, for energy thresholds~$\varepsilon_T = \sigma$, $2\sigma$, $4\sigma$, and~$8\sigma$. Tracers are selected as in Fig.~\ref{fig:ntpa_correlations} (injected after reconnection onset, above threshold at one light-crossing time after onset, and assigned by the plurality rule of Section~\ref{sec:shares}). Error bars show the statistical uncertainty (one standard deviation) from the finite number of selected tracers. Points with fewer than~$100$ selected tracers are omitted. The NTPA correlation at the three smallest boxes is nearly box-independent at the lowest threshold, while each doubling beyond~$L_x/d_e = 800$ lowers it.
    }
    \label{fig:ntpa_boxsize}
\end{figure}

Box size also affects the high-energy census. For~$\sigma = 50$, at~$\varepsilon_T = \sigma$ the unmagnetized NTPA correlation is nearly box-independent for~$L_x/d_e = 200$--$800$, and then falls monotonically out to the largest box. Figure~\ref{fig:ntpa_boxsize} isolates this box-size dependence of the unmagnetized NTPA correlation at four energy thresholds. Each doubling beyond~$L_x/d_e = 800$ lowers the unmagnetized NTPA correlation by~$\sim 20$--$45\%$ of its value at every threshold with at least~$100$ selected tracers, with the largest fractional drop occurring for the largest boxes. At the stricter threshold~$\varepsilon_T = 4\sigma$, the magnetized NTPA correlation overtakes the unmagnetized one, which falls fourfold from~$L_x/d_e = 400$ to~$6400$. The~$L_x/d_e = 200$ box holds fewer than~$100$ tracers above~$\varepsilon_T = 4\sigma$, and at~$\varepsilon_T = 2\sigma$ its unmagnetized NTPA correlation falls below that of the~$L_x/d_e = 400$ box although the two boxes tie at the lowest threshold. This pattern is consistent with the box size itself limiting the high-energy census, so the~$L_x/d_e = 200$ box constrains only the lowest threshold, and Fig.~\ref{fig:ntpa_boxsize} omits it above that threshold. Because the spectral cutoff grows with box size \citep{French2026}, part of the fixed-threshold decline could reflect thresholds sinking deeper below the cutoff. With thresholds rescaled by the box-size convention of \citet{Gupta2025}, $\varepsilon_T/\sigma \propto (L_x/d_e)^{1/2}$, the unmagnetized NTPA correlation still falls with box size. For the rescaled threshold that equals~$4\sigma$ at~$L_x/d_e = 1600$, and so runs from~$\sim 2.8\sigma$ at~$L_x/d_e = 800$ to~$8\sigma$ at~$6400$, the unmagnetized NTPA correlation roughly halves over that range. The apparent dominance of particles injected while unmagnetized in high-energy censuses may therefore be an artifact both of the finite temporal domain and of the finite spatial domain in kinetic simulations. Within a finite temporal domain, the channel whose tracers gain energy faster after injection is over-represented above a fixed threshold, whatever the mechanism behind the faster gain.

\section{Discussion} \label{sec:discussion}

In this work, we tested the hypothesis that $E>B$ regions are essential for injecting particles into the nonthermal population during relativistic magnetic reconnection at vanishing guide field \citep{Sironi2022, Totorica2023, Gupta2025}. The direct hypothesis (Section~\ref{sec:hypotheses}) holds that work delivered inside $E>B$ regions supplies most of the injection energy, and the indirect hypothesis asserts that energization during the re-magnetization just after an $E>B$ exit supplies most of it. To test both, we partitioned the pre-injection energy gain of a large tracer ensemble into work done while unmagnetized, while re-magnetizing, and while magnetized, in two-dimensional pair-plasma simulations across~$\sigma = 12$--$800$. This partition is generous to the $E>B$ hypotheses, since the re-magnetizing channel is credited with all work done within one gyration of an $E>B$ exit whether or not the encounter caused that work. By injection share, the magnetized channel is the largest contributor at every magnetization and under both plurality and pure assignment. The unmagnetized plurality share reaches at most~$\sim 20\%$, at~$\sigma = 800$. At~$\sigma = 50$, where we varied the box size, that share falls as the box grows, which suggests that finite boxes inflate it at other magnetizations as well. The fraction of injected tracers that never enter an~$E>B$ region before injection falls from~$\sim 90\%$ at~$\sigma = 12$ to~$\sim 40\%$ at~$\sigma = 200$--$800$. By high-energy census, particles injected directly by~$E>B$ energization are over-represented relative to their injection share, but nevertheless comprise a minority of the population above~$\varepsilon_T = \sigma$, and at~$\sigma = 50$ their NTPA correlation falls with box size beyond~$L_x/d_e = 800$. The injection shares reject both $E>B$ hypotheses of Section~\ref{sec:hypotheses}, since neither~$E>B$-associated channel is the largest contributor for a majority of injected tracers. Separately, injection without pre-injection~$E>B$ contact shows that such an encounter is not required for every injected tracer in these simulations. Below we compare these results with previous work, state their limits, and consider their implications.

\subsection{Comparison with previous work} \label{sec:comparison}

The magnetized channel is consistent with a known family of mechanisms. While a particle is magnetized, the electric field it samples is predominantly the motional field~$\mathbfit{E}_m = -\mathbfit{u} \times \mathbfit{B}/c$ of the bulk flows, which drives Fermi reflection and pickup acceleration \citep{Drake2006,Drake2009,Guo2014,Guo2015,French2023}. At weak but finite guide field, these mechanisms have been shown to supply the majority of the injection share \citep{French2023}, and in three dimensions at~$b_g = 0.3$ Fermi reflection is the dominant injection mechanism \citep{French2026}. The fact that the magnetized channel holds the largest injection share at~$b_g = 0$ is consistent with these motional-field mechanisms. In a pair-plasma simulation at~$b_g \approx 0$ and~$\sigma \approx 50$, \citet{Guo2025review} also find that most of the energy gained before injection by particles that cross $E>B$ regions is gained outside those regions. Our field-magnitude partition does not separately measure the work of these mechanisms or decompose the electric field into motional and nonideal parts.

\citet{Sironi2022} reported that $\gtrsim 90\%$ of the particles reaching $\gamma > \sigma$ at $\sigma \geq 50$ had passed through an $E>B$ region, and interpreted this result as evidence that nonideal fields solve the injection problem. That count includes contact at any time in a particle's history, including after injection \citep{Guo2023comment,Guo2025review}. Over the whole run, $\sim 70$--$85\%$ of our tracers above~$\varepsilon_T = \sigma$ at~$\sigma \geq 50$ enter an $E>B$ region at least once, whereas~$\sim 55$--$75\%$ enter one before injection. Neither contact statistic measures how much pre-injection work occurred inside~$E>B$ regions. Our census instead assigns tracers to channels by their pre-injection work fractions. At~$\varepsilon_T = \sigma$, tracers assigned to the unmagnetized channel make up at most~$\sim35\%$ of the census ($\sigma=200$ and~$800$), and the combined $E>B$-associated fraction, tracers with~$\xi_i^{\rm unmag}+\xi_i^{\rm remag}>1/2$, peaks at~$\sim60\%$. Read as evidence of injection, as in \citet{Sironi2022}, a contact census credits each $E>B$ encounter with all the energy a particle gains afterward. Our accounting reproduces this reading in the limit~$\phi\to\infty$, where a post-exit interval ends only at the next $E>B$ entry or at injection. Computed from each tracer's energy at its first $E>B$ contact, that accounting assigns the majority of the pre-injection gain to the two $E>B$-associated channels for $\sim 40$--$60\%$ of injected tracers at~$\sigma \geq 50$ ($\sim 8\%$ at $\sigma = 12$). These tracers make up~$\sim 85$--$95\%$ of those that enter $E>B$ before injection at~$\sigma \geq 50$, so in this limit nearly every tracer that touches an $E>B$ region is credited to an $E>B$-associated channel. The~$\phi\to\infty$ attribution therefore measures contact rather than the energy that contact supplies, and because it assumes that encounters cause injection, it cannot test that they do. Every tested finite amount of post-exit accumulated gyrophase,~$\phi = \pi/2$--$8\pi$, instead allows the magnetized channel to maintain the largest plurality share at every magnetization (Section~\ref{sec:shares}). At~$\sigma=50$, the~$\sim50\%$ of injected tracers that never enter~$E>B$ before injection is consistent with the~$\sim46\%$ reported by \citet{Guo2023comment} at this magnetization. The unmagnetized NTPA correlation also depends on the box size. Figure~\ref{fig:ntpa_boxsize} shows it falling at the $\varepsilon_T = \sigma$ threshold with each box doubling beyond $L_x/d_e = 800$ through the largest box we ran (Section~\ref{sec:ntpa}), so censuses measured at these sizes may overstate what a larger system would give. The re-magnetizing NTPA correlation instead falls far less over the same doublings (Fig.~\ref{fig:ntpa_correlations}(a)).

\citet{Gupta2025} define~$\zeta$ as the mean work done inside~$E>B$ up to the first kinetic-energy crossing of $\varepsilon^*m_ec^2$ where $\varepsilon^*=\sigma/4$, normalized by $\varepsilon^*m_ec^2$, for particles in a bin of kinetic energy~$\varepsilon_T$ at their evaluation time. In their usage, $\varepsilon_T$ thus labels a bin, whereas in ours (Section~\ref{sec:ntpa}) it is a threshold. At~$\sigma=50$, they report~$\zeta \simeq 0.4$ near~$\varepsilon_T/\sigma=2$ and~$\zeta\simeq0.8$ at~$\varepsilon_T/\sigma \gtrsim 8$. For our~$\sigma=50$, $L_x/d_e=1600$ run, the mean~$\xi_i^{\rm unmag}$ is~$0.16$ and~$0.27$ in factor-of-two bins of kinetic energy centered on~$2\sigma\,m_ec^2$ and~$8\sigma\,m_ec^2$, respectively, one light-crossing time after onset. Our binned means are lower than their~$\zeta$ by a factor of~$\sim 2.5$--$3$. The comparison establishes this difference without isolating its cause. Their quoted values use cold upstream plasma and time-averaged outflow simulations, whereas ours use the warm, periodic setup and population cut of Section~\ref{sec:methods}. The injection shares at~$\sigma = 50$ stabilize within~$\sim 0.25\,L_x/c$ of onset (Fig.~\ref{fig:shares}(a)), before the outflows can reach the periodic boundary, so we expect an outflow boundary to change the shares less than it would change the high-energy census, which is taken later. \citet{Gupta2025} report that~$\zeta$ falls with box size at fixed energy threshold and becomes box-independent once thresholds are rescaled to the maximum particle energy, which grows as~$L_x^{1/2}$. The highest tracer energies in our box-size scan grow as $L_x^{0.49}$, matching that scaling. Rescaling our thresholds to follow it weakens but does not remove the decline of the unmagnetized NTPA correlation for the threshold that equals~$4\sigma$ at~$L_x/d_e = 1600$ (Section~\ref{sec:ntpa}). The two scaling results, however, concern different statistics. Ours is the fraction of selected tracers assigned to a channel, whereas~$\zeta$ is a mean work fraction.

Our partition reads the energy budget of unperturbed trajectories. The test-particle experiments of \citet{Sironi2022} and \citet{Gupta2025} instead alter particle energies inside~$E>B$ regions. \citet{Sironi2022} resets the kinetic energy to a prescribed low value, which can also erase energy previously gained outside those regions \citep{Guo2023comment}. \citet{Gupta2025} preserve the momentum magnitude from the preceding step while allowing its direction to change. This procedure suppresses energization inside~$E>B$ without applying the same low-energy reset.

\citet{French2023} and \citet{French2026} measured injection shares under a mechanism partition that attributes each injected particle to acceleration by the electric field parallel to the magnetic field (their ``direct'' acceleration), to Fermi reflection, or to pickup. At $\sigma = 50$ and $b_g = 0.1$, the two motional-field mechanisms together inject $\sim 80\%$ of particles \citep{French2023}. At $b_g = 0.3$, parallel-field acceleration injects $\sim 50\%$ of particles in two dimensions ($\sigma = 8$--$96$) and $\sim 40\%$ in three ($\sigma = 8$--$32$) \citep{French2026}. Under our field-magnitude partition at $b_g = 0$, the magnetized channel is the largest contributor for most injected tracers at every magnetization. At $\sigma = 50$ its share is $\sim 70\%$, close to the motional share of \citet{French2023} at the same magnetization. \citet{French2026} also measured NTPA correlations. At $\sigma = 12$, particles injected by parallel-field acceleration make up $\sim 90\%$ of those near the spectral cutoff in two dimensions and $\sim 75\%$ in three, well above their injection shares. The over-representation appears at $b_g = 0$ as well, where the unmagnetized channel's NTPA correlation at $\varepsilon_T = \sigma$ sits well above its injection share at $\sigma = 200$--$800$. This excess above injection share persists at fixed injection time and comes from post-injection acceleration that differs by channel (Section~\ref{sec:ntpa}).

The re-magnetizing plurality share peaks at~$\sigma=50$ and then declines, while the unmagnetized share grows (Table~\ref{tab:channel_populations}). Between~$\sigma = 50$ and~$800$, the re-magnetizing NTPA correlation falls steeply (Section~\ref{sec:ntpa}), and at the stricter threshold~$\varepsilon_T = 4\sigma$ it falls from~$\sim40\%$ to~$\sim2\%$. At~$\sigma=800$, nearly all post-exit intervals end at re-entry before one gyration completes (Section~\ref{sec:shares}). This statistic measures interruption of these intervals, rather than the recovery or loss of adiabatic motion. The trends constrain the work credited to the post-exit interval, but do not establish whether the preceding encounter enables later acceleration for particles that enter~$E>B$. We have not identified the mechanism that energizes particles during post-exit intervals, and attributing post-exit work to $E>B$ regions would require identifying it. One candidate is a pickup-like process. In pickup, a particle that enters the outflow nonadiabatically is swept up by the flow and gains energy from the flow's motional field~$\mathbfit{E}_m$ \citep{Drake2009,French2023}. A particle that leaves an $E>B$ region into the outflow could gain energy in the same way while it re-magnetizes. Such work would be done by the motional field, although our partition credits it to an $E>B$-associated channel. By injection share, post-exit work is still not the largest contributor for most injected tracers at any tested amount of post-exit accumulated gyrophase~$\phi = \pi/2$--$8\pi$.

The review of \citet{Sironi2025} states that whether $E>B$ regions provide most of the energization during the injection stage ``remains debated'' and that broadly defined nonideal fields ``appear to be essential'' for injection. In our simulations, however, work done while particles are magnetized, outside any $E>B$ region and any post-exit interval, is the largest contributor to the injection energy for most injected tracers at every magnetization in the scan. The broadly defined form of the claim invokes the exact nonideal field and is a separate hypothesis to be discussed in Section~\ref{sec:limitations}.

\subsection{Scope and limits} \label{sec:limitations}

This study tested the role of $E>B$ regions in two-dimensional pair-plasma reconnection, while the broader question concerns nonideal electric fields in three-dimensional pair and electron-ion plasmas. Published $E>B$-dominance claims rest almost entirely on two-dimensional pair-plasma simulations \citep{Sironi2022, Totorica2023, Gupta2025}, the same setting as this study. The broadened claim of \citet{Totorica2023} is that the exact nonideal field, which has structure outside $E>B$ regions, dominates pre-threshold energization. No $E>B$-dominance claim has been advanced for electron-ion plasma. In three dimensions at~$b_g = 0.3$, Fermi reflection replaces parallel-field acceleration as the mechanism with the largest injection share \citep{French2026}, and a three-dimensional test at~$b_g = 0$ would be a separate study.

\subsection{Astrophysical implications} \label{sec:implications}

In the two-dimensional, nonradiative pair-plasma regime studied here, at vanishing guide field and~$\sigma=12$--$800$, injection can proceed without~$E>B$ contact, and the magnetized channel supplies the largest pre-injection contribution for most tracers. These results support energetic-particle models that allow injection outside~$E>B$ regions. Macroscopic models already represent acceleration by large-scale flows \citep{Guo2019, Li2018b, Arnold2021, Seo2024, Yin2024b, Li2025, Hu2026}. Applying such models to pair-plasma sources, including pulsar wind nebulae and magnetically dominated jets, requires establishing whether they reproduce injection rates and particle spectra in the relevant regime.

\section*{Acknowledgements}

O.F. acknowledges support by the National Science Foundation Graduate Research Fellowship under Grant No. DGE 2040434. F.G. acknowledges NSF under grant 2308091 and NASA under grant NNH240B72A. This research used resources of the National Energy Research Scientific Computing Center (NERSC), a U.S. Department of Energy Office of Science User Facility located at Lawrence Berkeley National Laboratory, operated under Contract No.~\mbox{DE-AC02-05CH11231} using NERSC award \mbox{FES-ERCAP0037511}.

\section*{Data Availability}

All simulations used the \code{VPIC} code \citep{Bowers2008}. The data underlying this article will be shared on reasonable request to the corresponding author.

\bibliographystyle{mnras}
\bibliography{main}

\bsp
\label{lastpage}
\end{document}